\documentclass[conference]{IEEEtran}
\IEEEoverridecommandlockouts

\usepackage{cite}
\usepackage{amsmath,amssymb,amsfonts}
\usepackage{algorithmic}
\usepackage{graphicx}
\usepackage{textcomp}
\usepackage{xcolor}
\usepackage{tabularx}
\usepackage{hyperref}
\def\BibTeX{{\rm B\kern-.05em{\sc i\kern-.025em b}\kern-.08em
    T\kern-.1667em\lower.7ex\hbox{E}\kern-.125emX}}
\begin{document}

\title{Improving Concept Learning Through Specialized Digital Fanzines\\
\thanks{This work has been partially funded by the Spanish Department of Science, Innovation and Universities: project RTI2018-099235-B-I00. It has been also partially funded by the project GR-2011-0040 from the University of Oviedo.}
}

\author{\IEEEauthorblockN{1\textsuperscript{st} Jose Manuel Redondo}
\IEEEauthorblockA{\textit{Computer Science Department} \\
\textit{University of Oviedo}\\
Oviedo, Spain \\
0000-0002-0939-0186}
}

\maketitle

\begin{abstract}
Specialized digital fanzines were successfully used to facilitate learning problematic concepts in an undergraduate programming course, dynamically adapting to student needs. The design of these fanzines favors creating and reading them quickly by establishing a common graphical layout, rules, and focusing in the most problematic parts of the concepts. This paper details the agile fanzine creation procedure, the way problematic concepts were identified and quickly handled, and how this approach was implemented in an actual course, so it could be applied to other courses with similar needs.
\end{abstract}


\section{Introduction}
\label{introduccion}

The \emph{Programming Technology and Paradigms} course in a \emph{Software Engineering} degree of a Spanish university \cite{EII} was designed to cover competences about the most common programming paradigms and its abstractions, including the basis of concurrent and parallel programming. This course uses a single programming language (C\#) due to time restrictions \cite{OrtinTPP}. The course covers a large range of concepts and, over the different course editions, multiple students declared that several concepts frequently turned more difficult to understand than expected. This causes frequent lack of synchronization with the course progression, and difficulties ``catching up'' with the current course lessons, due to the extra time used to properly understand these problematic concepts. The results of the course have been generally good, as most students could overcome these difficulties at the end, but complaints about the extra time spent in some parts were frequent. Thus, a way to improve student learning on these concepts was needed to try to put the expected average amount of self-study hours closer to the estimated one, as students undertake multiple courses in parallel and time should be adequately managed to optimize their academic results.

There are several works \cite{cervesato} \cite{Evans} \cite{Gonick} that use comic-like or fanzine-like approaches to teach concepts. Comics can be very complex, and their length is usually not restricted \cite{cervesato}. This way, they can be used to detail advanced concepts using a very practical approach \cite{Eisner85}. Fanzines are usually much shorter and quickly produced \cite{Sanders94}, and do not require a complex or lengthy edit process. Typically, fanzines are used in educational environments to make short introductions or approaches to certain concepts, as a prelude or complement to other materials. 
 
The main contribution of this work is the design procedure of a series of very specialized fanzines (abbreviated, zines) that have been used to improve the learning of problematic concepts identified by the students during an undergraduate programming course. Each zine focuses on clarifying concrete aspects of the problematic concepts rather than provide a full explanation, so they complement the rest of the course materials. The quick nature of fanzine creation is used to produce them on-demand, depending on students’ questions or requests. They are also thought to be read very quickly. This turned fanzines a very dynamic learning tool with a student-directed approach.  

This application of fanzines proved successful, as a great number of students declared that they helped to improve concept learning. Several students requested fanzines about specific concepts as the course progressed, giving them direct control of the process. This paper describes fanzine design rules and structure, how their topics were identified, its publication process, and user feedback and acceptance, so this technique could be applied or reproduced in other courses.
   
The structure of this paper is the following: Section \ref{related} describes the related work, while Section \ref{diseno} describes the course design and the concepts students found problematic. Section \ref{produccion} explains how fanzines were created using a three-phase creation plan, detailing rules, graphical layout, and all covered topics. Section \ref{resultados} describes the results and feedback obtained after using them and, finally, conclusions and future work are detailed on Section \ref{conclusiones}.

\section{Related Work}
\label{related}

Teaching with comic-like approaches has been used in all educational levels, from younger audiences \cite{Spiegel2013EngagingTW} \cite{Yu2015} to university-level courses. They are also not only applied in Computer Science, but also in other types of studies \cite{Yu2015}, such as Mathematics \cite{Reilly09}. Comics have also been used to try to explain substantially complex technical concepts to the general public, such as the inner workings of the \emph{Google Chrome} browser \cite{McCloud08}. A more modern example is \cite{Ramos18}, a comic book novel that uses a fictional but realistic history as a central way to explain multiple security concepts and practical applications. 

Using a comic-like approach to explain Computer Science concepts is not a new technique; there are examples from more than 30 years ago: \cite{Gonick} explains the history of computing and some Computer Science fundamentals, introducing concepts in a way that may appeal to most reader types. Although in our particular case we applied this research to programming, comic-like approaches have also been applied to more topics. For example, we have applications to explain computer networks \cite{Ganesh2013TheEO} that also use comic strips with rich contents as supplementary materials. Logic for Computer Science major courses has also benefited from this approach, using a comic book as the course textbook \cite{cervesato}. Multiple Computer Science topics are also covered in \cite{Yim09} \cite{Yim10} by using a series of comic illustrations as complementary class handouts to reinforce concept learning. As in our case, these illustrations are also combined with concise explanations, although not with a predefined fixed structure that accelerate their production like our approach.

One of the closest works related with our research successfully uses comics applied to programming courses but using an inverse approach \cite{Zaibon2018EnhancingPO}. In this research, a student control group is instructed to produce comics that reflect their understanding of certain selected course topics with the \emph{BitStrips} authoring tool. This way, students reinforce the understanding of the concepts by trying to explain them through a comic, instead of being the teacher the one that delivers explanations via comics as in our approach.

Outside regulated courses, fanzines are also used as educational tools in different contexts. One of the closest works to the one described in this paper is \emph{Wizard Zines} from Julia Evans \cite{Evans}. These zines focus on introducing a variety of different concepts on multiple fields of Computer Science, rather than explaining specific conceptual problems identified by a concrete audience. Julia has published books with compilations of her fanzines grouped by topic, such as \emph{Git}, and \emph{Linux} command-line tools and concepts. As in this research, these zines follow a coherent visual style and divide their contents in frames, although \emph{Wizard Zines} use simpler visuals and are much more focused on text-based explanations. The main difference between \emph{Wizard Zines} and the research described in this paper is that this one is directed by the audience needs: it focuses on the problems detected in a concrete course. New material creation is driven by student needs or requests while the course is taking place, rather than introducing computer science concepts to a general audience with different skills with no special time restriction. 

Another very similar approach to \emph{Wizard Zines} is the \textit{BubbleSort Fanzines} \emph{Kickstarter} project of Amy Wibowo \cite{Wibowo}. Her aim is to make a variety of computing tools, skills, and concepts accessible to everyone. Fanzines are used here to reframe concepts, so they could be found attractive by audiences that initially did not consider studying them or enroll in certain courses. Planned zine topics are very varied, such as logic circuits, hardware hacking, cryptography, the basis of Internet communication, and some programming concepts such as recursion and sorting. Plans to develop zines about advanced topics such as neural networks, AI, image processing, and computer graphics also exist. This approach also focuses on text frames coupled with hand-made images of concepts, following a simple and clear style. Humorous titles are used to introduce each zine topic. 

Lin Clark also uses a comic-like approach to explain advanced concepts regarding networking, web servers, and web technology concepts \cite{Clark}. Although the topics covered by these comics are all focused on web technologies, they are independent, not following a concrete creation plan or line. This is the main difference with the research described in this paper, that develops fanzines on demand following a course syllabus. Regarding style, these comics are mostly composed by sort paragraphs of text providing explanations followed by an illustrative image (mostly simple line drawings, like \emph{Wizard Zines}). There is no apparent limit related to the amount of text or number of comic frames, as some comics use a very different number than others.

\emph{Xkcd} \cite{Munroe} are very popular technology-related comics dealing with a wide variety of topics: from technology-focused ones to others not related with computing. These comics emphasize the humoristic part to try to transmit a message to the readers, even ignoring technical parts that may be involved in the message, or not being rigorous explaining them. Comic size is very varied, from just one comic frame to multiple ones. Topics do not follow any structure, and comic frames are mostly composed by line drawings plus small texts. 

\emph{Little Bobby} \cite{Lee} is a long-running web comic strip series (250+ released comics), released weekly, and based on the comic book \cite{Lee13} of the same authors. The comic strip topics are varied including, but not limited to, SCADA, the cloud, net neutrality, hacking, and big data. It has a lighthearted nature, focusing on making technical and often difficult topics more accessible to a wider audience. It also uses humor and inside jokes for the topic’s respective community to achieve their goals. Comics usually are composed by full-color 3-4 frames, using main recognizable characters and short texts within speech bubbles between them as in the approach described in this paper. Characters are represented in a wide variety of situations and environments, and have widely different poses, as they are drawn by a professional illustrator. Although this kind of drawings makes each released comic substantially different from the others, this kind of complexity cannot be implemented in the context that will be described in this paper, as the author lacks the necessary drawing skills to develop them within the time constraints of the course. 

These implementations of comics and fanzines as learning tools prove that they can be used effectively to introduce a great amount of different Computer Science topics (among other disciplines), both basic and advanced, inside, or outside regulated courses, at any educational level, and facilitating their learning to a varied audience. However, the implementation described in this paper shows the benefits of following a more focused approach, using zines to reinforce topics complementing existing materials, and quickly developing them on-demand. Another benefit of this research is to detail a common development framework and rules to try to speed up fanzine releases during a course, so this approach can be reproduced in other contexts with the same goals.

\section{Course Design and Problematic Concepts}
\label{diseno}

The \emph{Programming Technologies and Paradigms} course was designed with five mandatory units \cite{OrtinTPP} plus optional materials. It follows the \emph{Programming Methodology} course, fully taught in Java. Our course also uses a single language due to course time restrictions (6 ECTS credits), but Java was judged unsuitable because its limitations in generics (due to its \emph{type erasure} implementation \cite{Bracha98}), inability to provide functions as first-class entities, and no support for certain key functional programming concepts (continuations, \emph{lazy} evaluation, \emph{pattern matching}, and \emph{comprehension lists}).
 
Due to Java limitations C\# was chosen instead, allowing students to gain skills in another very popular language, both in academic and professional contexts (\cite{TIOBE} places C\# as the 6th most used language). C\# has full support to basic and advanced object-oriented features, implements functions as first-class entities, \emph{lambda expressions}, \emph{closures}, \emph{comprehension lists}, and a form of \emph{continuations}. Additionally, the \emph{Language Integrated Query (Linq)} library is a very popular data manipulation framework that uses C\# functional programming features. Concurrency and parallelism are supported via asynchronous message passing, explicit thread creation, and the more advanced \texttt{Task} abstraction. Data and task parallelization can be simplified using the \textit{TPL} (\emph{Task Parallel Library}) and \textit{PLinq} (\emph{Parallel Language Integrated Query}) libraries \cite{Terrell18}. Typical synchronization mechanisms such as \texttt{lock}, \texttt{Mutex} and \texttt{ReaderWriterLockSlim} are also present. Additionally, \emph{metaprogramming} features (introspection, structural intercession, and dynamic code evaluation) and \emph{dynamic typing} (via the \texttt{dynamic} type) are supported. The following subsections detail what concepts were found problematic in each unit.

\subsection{The object-oriented programming paradigm}

This unit ``links'' with the previous course, also introducing new concepts like \emph{bounded generics} and \textit{type inference}. Surprisingly, a considerable number of students needed more time than expected to deal with concepts they should already know. Part of this problem could be due to the 9-month gap with the preceding course, as courses in-between in the official curricula do not facilitate acquiring or reinforcing programming skills (\emph{Computer Electronics Technology}, \emph{Computer Architecture}, \emph{Computability}), or deal with specialized applications (\emph{Data Structures}, \emph{Human-Computer Interaction}). Therefore, problematic concepts in this unit were a mix of ``old'' and ``new'' ones.

\begin{itemize}
\item	Parameter passing was weakly understood in general. The concept of object reference was not used correctly by a surprising number of students that treated basic and reference types the same, so the predicted outcome of some examples was consistently guessed wrong. This worsened when the \texttt{ref} and \texttt{out} keywords were explained.  
\item	Multiple students had trouble differencing between \texttt{Equals} and \texttt{==}.
\item	There were considerable problems with type conversions and usages of the \texttt{as} and \texttt{is} keywords.
\item	Multiple students could not properly differentiate acceptable usages of exceptions and assertions, understanding the consequences of using them. Therefore, design by contract was also challenging. 
\item	Given that C\#, unlike Java, do not enable dynamic binding by default, and a previous weak understanding of this mechanism, practical applications of polymorphism and dynamic binding were also a problem. 
\item	Dealing with generics was also difficult for some students, especially when bounded generics usage scenarios were introduced. Students were used to the restrictions imposed by the type erasure generics implementation technique of Java \cite{Bracha98}. C\# does not use this approach, so it does not have these restrictions, but students still behaved as if they were enforced.  
\item	The concept of iterator (implemented through \texttt{IEnumerable} in C\#) was poorly handled by most students, as some even tried to use them as full-featured data structures. Understanding iterators is key for the next unit, so reinforcing this concept was extremely important. Coordination with the \emph{Data Structures} course should also be improved to avoid this.  
\end{itemize}

\subsection{The functional programming paradigm}

At the end of this unit students must be able to design and implement applications using suitable elements provided by the functional paradigm, comparing functional and object-oriented approaches. This paradigm is new for the students and handling even the most basic concepts proved troublesome to several of them. 

\begin{itemize}
\item	Multiple students struggled understanding that functions are first-class types. Understanding how code can be stored in data structures, passed as parameters, or returned from functions, took substantially more time than expected to use effectively. 
\item	Creating and using code ad hoc via lambda functions is a powerful mechanism, but students initially avoided that as they failed to understand how to use it to their advantage. The foundations of lambda calculus, taught to explain the origin of the functional paradigm and lambda functions, also proved troublesome when the students were preparing the final course theory exam. 
\item	Another problematic concept was lazy evaluation. Students only had experience with languages implementing eager evaluation. This increased the difficulty of applying the limited implementation of this concept in C\# (generators) correctly, and its related concept, continuations.  
\item	Closures and partial application proved also challenging. The first, due to improper handling of free variables, as students generally did not understand that closures hold a reference to them, not a copy. The second, because they could not easily see the advantages of using currying to obtain new code by applying part of the function parameters. This concept was especially difficult, as currying functions to implement partial application must be done manually in C\#.
\item The final part of this unit reviews a popular framework, \emph{Linq}, to show how functional programming concepts are applied in a professional environment. Although basic higher-order functions (\emph{Map}, \emph{Filter}, and \emph{Reduce}; \texttt{Select}, \texttt{Where}, and \texttt{Aggregate} in C\#), were generally quite well understood, concatenating the output of a function with the input of a following one was rarely used at first. The lazy nature of these functions also was troublesome, especially why code passed as parameters did not execute unless elements are retrieved from the results of \emph{Linq} queries. Finally, more complex \emph{Linq} operations such as \texttt{Join} or \texttt{GroupBy} were also poorly handled due to confusion with the SQL operations with the same name, as the \emph{Databases} course is taught in parallel with this course. 
\end{itemize}

\subsection{Concurrent and parallel programming} 

Students must be able to know and apply the basic elements of concurrent and parallel programming at the end of this module. They demonstrated to have a good understanding of threads (taught in \textit{Programming Methodology} and used in the parallel \textit{Operating Systems} course), but demonstrated difficulties when using mechanisms to handle critical sections (especially the \texttt{lock} keyword and the more complex \texttt{ReaderWriterLockSlim} class). Effectively using typical thread-based schemas (such as \emph{Master-Worker} and \emph{Producer-Consumer}) proved also challenging to some students due to poor understanding of the relationships between the classes that compose them. Finally, several students also experienced problems when parallelizing existing single threaded code using the \emph{TPL} or \emph{PLinq} libraries. 

\subsection{Meta-programming and dynamic typing} 

Dynamically typed programming languages have influenced the development of software in the last years \cite{Ortin14} and are a very active research line in our Computer Science research group \cite{OrtinSCP}. This last unit identifies and discusses the distinguishing features they provide, such like \emph{multiple dispatch} \cite{Ortin14b}, \emph{dynamic typing} \cite{Lagartos19} or \emph{generative programming}. This unit is merely introductory, so the only important problem students had was understanding how some concepts (\emph{dynamic typing}, \emph{dynamic code evaluation}\dots) could be simulated in C\# to provide functionalities similar to those implemented by typical dynamically typed languages, such as \emph{Python} \cite{RedondoPython}. 
 
\subsection{Optional materials} 

These public materials allow willing students to further study the acquired concepts, even when the course is finished. They are also used to promote course enrollments and to provide updated materials to students that successfully took the course in the past. They include how to apply the functional features they learned in Java [16], and features of C\# 6, 7.X, and 8.X not explained during the course. The last 3 have accumulated 567, 115, and 163 reads respectively since January 2019 in \emph{ResearchGate}. The only questions were related to dynamic code generation through the \emph{Roslyn} Compiler-as-a-service feature and the usage of the \emph{.Net} \emph{Speech} APIs.

\section{Fanzine Creation}
 \label{produccion}

Although the previous section described the complete list of problematic concepts, each problem was not known outright, but discovered as the course was being delivered. Fanzines were used as answers to questions about these concepts as they were known, to reinforce their understanding on-demand. To do this, an action plan to produce fanzines before the course started was created in a way it could be applied to other courses if the approach proved successful. The action plan was divided in three phases: \emph{design of the fanzine framework}, \emph{lecturer-directed fanzine development}, and \emph{student-directed fanzine development}.

\subsection{Design of the fanzine framework} 
In order to be able to answer student needs using fanzines in an agile way, this initial phase established several rules to accelerate fanzine creation. All fanzines were thought as very short stories narrated by a main character. 

\begin{figure}[!t]
\centering
\includegraphics[width=\columnwidth]{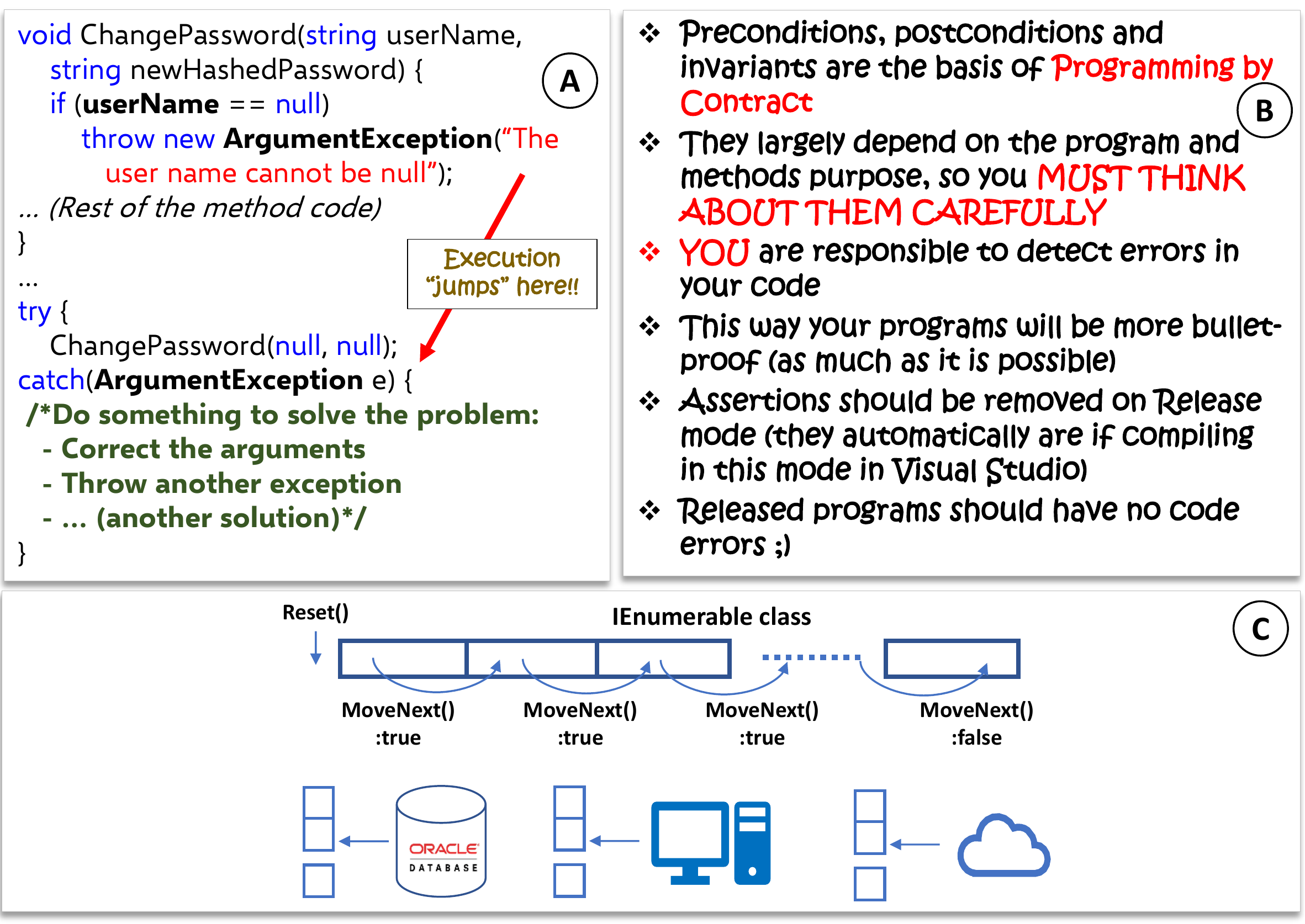}%
\caption{Different fanzine frame types}
\label{fig01}
\end{figure}

\begin{figure*}
\centering
\includegraphics[width=14cm]{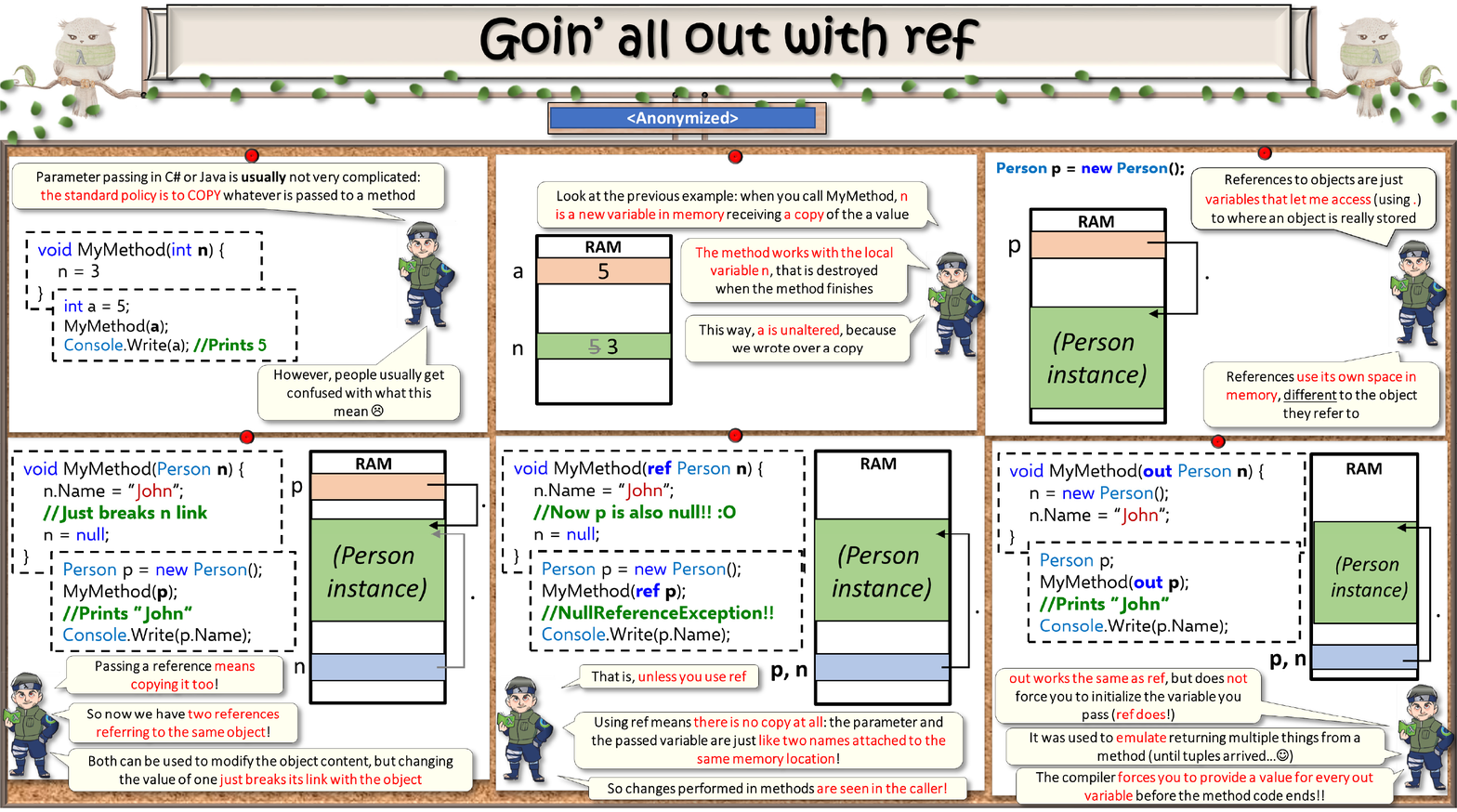}%
\caption{C\# parameter passing strategies fanzine}
\label{fig02}
\end{figure*}

\begin{figure*}
\centering
\includegraphics[width=14cm]{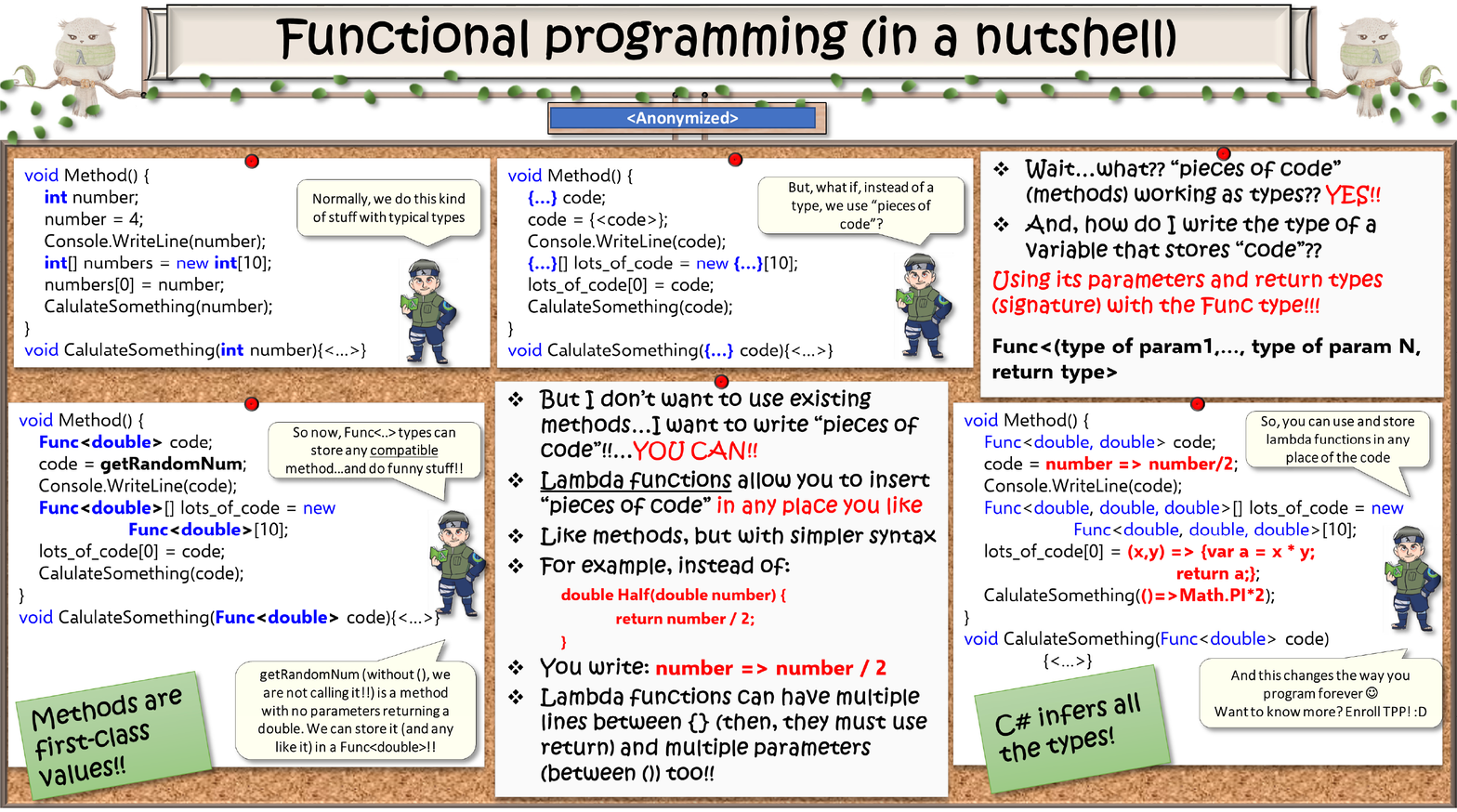}%
\caption{Basic functional programming principles fanzine}
\label{fig03}
\end{figure*}

\begin{enumerate}
\item	All fanzines were created using a default installation of \emph{Microsoft PowerPoint}, with no third-party plugins.
\item	Always use 6 comic frames (two rows of 3 frames) to give an appropriate amount of information to the reader, according to the findings of a classical work regarding this \cite{Miller56}, and to facilitate quick reading of their contents.
\item	A main character narrates the explanations on each comic frame. Widely known personalities related to computer engineering were considered, such as Alan Turing, Ada Lovelace, John Von Neumann, or Margaret Hamilton. However, their importance and contribution to computing was too important to relate them to a ``simple'' programming technology course. Therefore, to better relate fanzines with the course, a caricature of its head teacher was used. This main character is a single image (no different poses were created) to increase creation speed.  Main character explanations are provided as speech bubbles with a common format. 
\item	Only three types of frames are used to explain contents (see Figure \ref{fig01}): commented source code fragments (\emph{A}), text-only explanations in bullet lists (\emph{B}), and drawings/diagrams (optionally with source code snippets) (\emph{C}). This is to achieve a coherent explanation style among the different fanzines. 
\item	All fanzines are created on a single page (default horizontal \emph{PowerPoint} slide) to facilitate quick reading and sharing. This way, they can be easily distributed as images or quickly viewed in any device. As their goal is solving specific conceptual problems, contents are always very direct, focusing only in solving problems and trying not to give excessive details. This requires a special effort to carefully choose what it is included, as space is quite limited. Zines are thought as complementary resources, not replacements of the rest of the course materials that are responsible of explaining concepts in a much broader perspective. They must be read and understood quickly to be used as intended. 
\item	Fanzine frames follow the occidental reading order (left to right, upper to lower row), as it is the one used by all students. 
\item	Every fanzine has the same header, consisting in a wooden stand with a folded banner, leaves, and two decorative owl characters at each side (see Figure \ref{fig02} and Figure \ref{fig03}). They also have the same appearance: a wooden bulletin board with six notes fixed with a pushpin.
\item	Regarding content, all fanzines share the same colors and font features (typeface, size, line length, line spacing, character and word spacing, indents\dots) for titles and contents \cite{Itkonen}, to be recognizable as part of a group.
\item	Finally, as several of the related works described in Section \ref{related}, fanzines use light touches of humor to be more approachable and to reinforce its comic nature. These are mostly present in the fanzine title, to enable us to focus on contents later. 
\end{enumerate}

Following these rules, a base visual template was created to create each fanzine from. This way, new fanzines just need to add a title in the provided space and fill the frames with explanations following rule 4. Occasionally, sticky notes could highlight very important things to remember. Fanzines are published classified by course unit and identified by a short text describing its contents (see Table \ref{tabla1}), so they can be easily located. 

\subsection{Lecturer-directed fanzine development}

The goal of the second phase is to start fanzine development. Although the experience of previous course editions could be used to identify potential problematic concepts and start creating fanzines, only feedback from the current course edition was used. This allowed testing if this approach could be successfully developed during a typical course, and to dynamically adapt to the actual problems experienced by students, that might differ from previous editions.  

This way, a fanzine was created when several questions related to the same concept began to appear in the different communication channels the students have (email, private course forums, office hours, or even social networks). Thanks to the rules described in the previous section, zines were developed very quickly, as more complex ones took up to two hours to develop, always by a single lecturer. Once created, zines were made available using the course private virtual platform to handle course materials, so every student with the same problems may benefit from them. 

Finally, created fanzines were also made public each week. This proved useful to students that already took the course but wanted to refresh some contents (as several were using them in their professional activities), to obtain feedback, and to promote future course enrollments. 

\subsection{Student-directed fanzine development}

The third stage of our action plan aimed to give students some degree of direct control of the fanzine creation process, once the second phase was active during the first course unit (object-oriented programming). At the beginning of the functional programming unit, students were informed that they can directly request specific fanzines about any of the upcoming or past course topics. When several requests related to the same concepts were received, the corresponding fanzine was developed following the described rules and publication procedure. Fanzines created during this phase cannot be visually distinguished from those created in the previous one. Choosing the second course unit as the starting point was motivated by the possibility of students having more conceptual problems, as most of the concepts in the course will be new for them from there on. Phase 3 runs in parallel with phase 2, so both can be used to determine the new fanzines to be created.

This third phase not only allows students to decide what new fanzines are going to be created: they are also welcomed to give feedback about any existing fanzine or ways to improve the whole initiative. Apart from some minor errata, substantial feedback and several interesting requests were received during the course and, once considered, quickly incorporated to already created fanzines thanks to their design rules. 

The most surprising feedback was multiple requests for Spanish translations. This course is delivered both in English and Spanish, and students freely choose which group they want to be in. It was assumed that the average English level of the Spanish group students was good enough to be able to read small pieces of text. Therefore, all fanzines were initially created in English to reach a potentially greater audience. However, several students of the Spanish group reported that it was substantially more difficult to correctly understand the explanations in English, which is contrary to the goals of this research. Fortunately, the nature of fanzines also facilitated their fast translation. These translations include source code (variable and method names, strings…) and certain expressions. When translation was a clear student priority, all subsequent zines were directly produced in both languages and published in separate sections. 

Format problems were also reported. The \emph{Kristen ITC} font was used to emphasize the comic appearance. However, several students complained that it made source code more difficult to read due to the font variable size and closeness of the text paragraphs. To solve this, an extra separation was introduced between text paragraphs, and a more console-like font with no variable font height was used for source code snippets (\emph{Selawik}), both in existing fanzines and in the base template for new ones. Additionally, some students indicated that the usage of the red color to emphasize key concepts inside explanations was excessive in some of the fanzines, so its usage was minimized. Figures in this paper incorporate these format changes.

Finally, several students complained about clearness and resolution problems when viewing zines in portable devices. This happened because they were initially provided as \texttt{.png} images with sub-HD resolution to minimize their size and thus facilitate sharing them. However, zooming over its contents leave indeed a poor result, so they were released again individually in high resolution PDF format. 

\subsection{Distribution of created fanzines}
\label{distribucion}

29 fanzines were developed to address all the problems described in the previous section (26 during the 2018 course year and 3 more in the following year). Table \ref{tabla1} shows their information classified by course unit, showing its topic and types of frames used (see Figure \ref{fig01}). This table also specifies (\emph{Ph} column) if a fanzine was created during phase 2 (lecturer-directed) or phase 3 (student-directed).

Figure \ref{fig02} and Figure \ref{fig03} show two fanzine examples. Fanzine in Figure \ref{fig02} explains the parameter passing (\texttt{ref} and \texttt{out}) concept belonging to the object-oriented programming unit, addressing the problems outlined in the previous section. Likewise, Figure \ref{fig03} does the same with the basics of functional programming of the functional programming paradigm unit. Both belong to the fanzine list contained on Table \ref{tabla1}. 

Table \ref{tabla1} shows that 9 fanzines were developed to reinforce the new and known concepts of the object-oriented paradigm; 10 were dedicated to reinforcing functional programming concepts, and 7 were created to reinforce concurrency and parallelism concepts. Finally, 3 more were created to understand applications of the final metaprogramming unit and optional materials. Additionally, 13 zines were developed by direct student request (phase 3) and 16 developed from the students most frequent questions (phase 2). Fanzines requested by students concentrate at the end, as students seemed more engaged with this initiative as the course progressed. A final compilation of all zines was also made available in a public site at the end of the course \cite{RedondoZine20} as a high-resolution PDF file (one for each language). Additionally, a table with a detailed description of each zine contents and the specific problems they try to address is available in \url{https://www.researchgate.net/publication/348662668_Detailed_description_of_each_zine_contents}. All fanzines were developed as a consequence of the problems highlighted in Section \ref{diseno}.

\begin{table}[!t]
\renewcommand{\arraystretch}{1.3}
\caption{Fanzines per course unit}
\label{tabla1}
\begin{tabularx}{\columnwidth}{Xcc}
\textbf{Contents} 	& \textbf{Frame types} & \textbf{Ph} \\

\hline
\multicolumn{3}{c}{\textit{Object-Oriented Programming Unit}} \\
\hline

Different behavior of dynamic binding in C\# and Java & A, B, C & 2 \\
\texttt{Object.Equals}, equality redefinition and \texttt{==} operator & A, B & 3 \\
Type conversions, \texttt{is} and \texttt{as} & A, B & 3 \\
By value and by reference parameter passing, \texttt{ref} and \texttt{out} & A, C & 2 \\
Exceptions and preconditions & A, B & 2 \\
Asserts and postconditions & A, B & 2 \\
C\# generics, type variables & A, B & 2 \\
C\# generics vs Java generics & A, B & 3 \\
How \texttt{IEnumerable} works & B, C & 2 \\

\hline
\multicolumn{3}{c}{\textit{Functional Programming Unit}}  \\
\hline

Functions as first-class objects & A, B & 2 \\
Using predefined delegates & A & 2 \\
From method to lambda expression & A & 2 \\
Basics of \textit{Linq} and \texttt{IEnumerable} manipulation & A, B, C & 2 \\
Generators and \texttt{yield return} & B, C & 2 \\
Lazy policy of \textit{Linq} functions & A & 2 \\
Alpha-conversions and beta-reductions in lambda calculus & C & 3 \\
How \textit{Linq} \texttt{Join} function works  & A, C & 3 \\
How \textit{Linq} \texttt{GroupBy} function works & A, C & 3 \\
Currying a function to achieve partial application & A, B & 3 \\

\hline
\multicolumn{3}{c}{\textit{Concurrency and Parallelism Unit}} \\
\hline

\texttt{lock} and critical sections; thread-safe data structures & A & 2 \\
Closures on threads, critical section problems & A & 3 \\
\texttt{lock} usage for critical section protection & A, C & 3 \\
Using \texttt{ReaderWriterLockSlim}  & A & 3 \\
\textit{Master-Worker} schema & A, B, C & 2 \\
\textit{Producer-Consumer} schema & A, B, C & 3 \\
\textit{TPL} and \textit{PLinq} to parallelize code & A & 2 \\

\hline
\multicolumn{3}{c}{\textit{Metaprogramming and optional units}}  \\
\hline

Emulation of runtime intercession & A, B & 2 \\
\textit{Roslyn} to enable dynamic code generation & A, B, C & 3  \\
Text to speech conversions & A & 3  \\
\hline
\end{tabularx}
\end{table}

\section{Results and Discussion}
\label{resultados}

The goal of fanzines was not to try to improve the overall course results (see Section \ref{introduccion}), as they are already adequate (see Table \ref{tabla2}), but to decrease the time students use to understand the most problematic concepts. Zines might even have a negative impact in the course results if they are used as its main study materials. This is because they were not designed to explain full concepts, just to reinforce their most problematic parts. Using them as the only reference material to study concepts may lead to understanding them incompletely. To check if this effect is produced, four course editions were analyzed, only the last using zines. Table \ref{tabla2} shows that fanzines did not have a negative impact on course results, being roughly the same as in the previous edition.

\begin{table}[!t]
\renewcommand{\arraystretch}{1.3}
\caption{Course results}
\label{tabla2}
\begin{tabularx}{\columnwidth}{Xcccc}
\hline
										  & \textbf{15-16}  & \textbf{16-17}  & \textbf{17-18}  & \textbf{18-19}  \\
\hline
Number of students & 153    & 185    & 147    & 140    \\
Average mark    & 5,16   & 6,39   & 6,95   & 6,83   \\
Standard deviation   & 2,4    & 1,97   & 1,67   & 1,76   \\
Minimum score     & 0,38   & 0      & 2,5    & 1,57   \\
Maximum score     & 9,53   & 9,85   & 10     & 10     \\
\% pass rate  & 56,9\% & 63,8\% & 75,5\% & 75,7\% \\
\hline
\end{tabularx}
\end{table}

Unfortunately, systematically evaluating the time saved understanding concepts by using fanzines was considered impractical, as no experiment could be thought to accurately measure it in this context. The time students spend understanding different concepts depends on their individual abilities, previous knowledge, and environment (available help from other sources). Also, this research is also created to be applied on-demand during a course, so typical systematic evaluation techniques, like creating control groups or applying use case studies \cite{Wohlin12}, could not be designed without disrupting the normal course flow, forcing students to use extra time to fulfill these tasks, which is contrary to the goals of this research. Thus, the evaluation approach is based on student personal opinions about the perceived usefulness of fanzines on their own case. This approach might introduce some threats to the validity of the results, but it is the only one that could be applied in this context without causing problems.

Therefore, to evaluate the usefulness of zines perceived by their users, feedback from different user types was collected once the first 8 were released. Feedback recollection was designed to be quick, optional, and not disruptive with the course flow. The first feedback collection process targeted to any potential user, no matter if it is a student of the current course edition or not. The head teacher private \emph{Twitter} account was used to create polls. This account had a restricted set of 370 followers, mostly composed by old course students, computer engineers from other universities, or computer professionals of different fields, manually chosen from follow requests to avoid robots, fake accounts, and people with no verifiable background. As \emph{Twitter} polls are anonymous and very restricted, only two questions were made: 

\begin{enumerate}
\item	``Do you think that, in general, fanzines help understanding concepts?'': 46 answers were given, 93\% positive. 
\item	``Do you want a compilation of all fanzines available to the public?'': 111 answers were given, 96\% positive. 
\end{enumerate}

Results show that zines are perceived as useful for their intended purpose by most users, and that there are a substantial number of persons interested in obtaining a compilation of them. This is the main reason why the zine compilation \cite{RedondoZine20} was released at the end of the course (see Section \ref{distribucion}). 

Obtaining feedback from current course students was the next step in the feedback collection process. To do that, the private online course site was used to create a more complex anonymous poll. This poll had 5 questions with multiple answers (see Table \ref{tabla3}), and a final free text one (see Table \ref{tabla4}). For questions with N $>$ 2 answers, scores from 1 to N were assigned, and the arithmetic mean was calculated. 62 answers were collected (44.3\% of the enrolled students). Free text answers also carried a lot of feedback. Table \ref{tabla4} classifies the contents of free text answers into categories to try to extract useful information, considering that one answer may contain information falling into several categories. These free-text answers were also used to determine new zines to be created during phase 3. Additionally, a parallel poll was also created through \emph{Google Forms} to obtain the same feedback from any zine user, using the same questions except the second. 44 users answered this poll. In both cases users can only answer once.  

Table \ref{tabla3} shows that most students (96\%) and users (97,7\%) found zines useful or partially useful to learn concepts. However, the selection of created fanzines had room of improvement, as 21\% of the students were looking for fanzines about topics not already covered. This percentage started in roughly 30\%, but lowered as more fanzines were released, indicating that on-demand fanzine creation had a positive effect during the course. It is reasonable to assume that most students answering \emph{Yes} in this question also used the free-text question to ask for the concrete fanzines they wanted (see Table \ref{tabla4}).  Fanzine content design is considered very satisfactory: no student declared problems with their length (only 7\% of general users do), and 95\% agree with the way they cover concepts, so the quick reading objective was met. 11.6\% of the general users consider that they contain too much information, opposite to only 3\% of the students, indicating that their contents are viewed as specialized complements of the existing materials.  

\begin{table}[!t]
\renewcommand{\arraystretch}{1.3}
\caption{Zine feedback polls results}
\label{tabla3}
\begin{tabularx}{\columnwidth}{XXX}
\hline
\textbf{Question} & \textbf{Result (Students)} & \textbf{Results (General Users)} \\
\hline
Did fanzines help you to learn concepts?  & 85\% Yes \newline 11\% Partially \newline 3\% No \newline \textbf{Mean}: 2,8 / 3  
& 93\% Yes \newline  4,7\% Partially \newline  2,3\% No \newline \textbf{Mean}: 2,9 / 3 \\
\hline

Did you miss fanzines about concepts that you found complex?  & 21\% Yes \newline  79\%No & - \\
\hline

Do fanzines highlight the correct parts of the concepts?   & 95\% Yes \newline  3\% Too much data  \newline 2\% Missed key parts  \newline 0\% No \newline \textbf{Mean}: 3,9 / 4
& 86\% Yes \newline  11,6\% Too much data  \newline  0\% Missed key parts   \newline 2,4\% No \newline  
 \textbf{Mean}: 3,8 / 4  \\

\hline

Do you think that fanzines have an adequate length?  & 100\% Yes \newline  0\% No & 93\% Yes \newline  7\% No \\
\hline

Do you think that fanzines like this could be useful in other courses? & 93\% Yes \newline  7\% Programming only  \newline 0\% No  \newline \textbf{Mean}: 2,9 / 3  
& 93\% Yes \newline 4,7\% Programming only \newline 2,3\% No \newline \textbf{Mean}: 2,9 / 3 \\
\hline

\end{tabularx}
\end{table}

\begin{table}[!t]
\renewcommand{\arraystretch}{1.3}
\caption{Free text poll feedback}
\label{tabla4}
\begin{tabularx}{\columnwidth}{Xcc}
\hline
\textbf{Feedback category} & \textbf{Students} & \textbf{General} \\
\hline
No comments  & 31,5\% & 58,14\% \\
Explicitly like the zine approach  & 42,6\% & 23,81\% \\
More zines or apply them in other courses  & 21,3\% & 14,29\% \\
Request format changes  & 13,11\% & 4,76\% \\
Request content changes  & 6,56\% & 4,76\% \\
Request translation  & 11,48\% & 0\% \\  
\hline                     
\end{tabularx}
\end{table}

Finally, most students and users (93\%) consider that this approach could be successfully used in other courses. On the other hand, Table \ref{tabla4} shows that a substantial number of students, and more than 50\% of other users, do not give any comment. 42.6\% of the students explicitly declare that they liked the idea (23.81\% in case of the users). New specific zine requests, or an explicit request to use them in other courses, was given by 21.3\% of the students (14.29\% in case of the general users). Concrete format and content changes were requested by few people, and Spanish translation was only requested by students. 

Finally, our university uses anonymous polls over each course and lecturer to collect general student feedback. These mention zines in some of the comments, stating that they have been considered a great improvement of the lecturer’s support to student problems. The author of the paper was also evaluated with an overall 9.8/10 average score in the \emph{teaching}, \emph{attitude}, and \emph{satisfaction with the work} poll categories.  

\section{Conclusions and Future Work} 
\label{conclusiones}

This paper describes how specialized fanzines have been successfully introduced to reinforce the learning of problematic concepts in an undergraduate programming course. The main contribution is the fanzine design procedure and application, based on explaining very concrete concepts, a common base template to fill with contents, and a static set of rules. This allows to respond quickly and dynamically to student needs, creating fanzines that adapt to the necessities of each course edition and that can be read quickly. Application of this technique have been received very positively both by students and other potential users, explicitly indicating that they reached their intended objective. Users also consider that this approach could be applicable to other courses, even not related to programing. 

The zine compilation released to the public was published in May 2019 \cite{RedondoZine20} and has accumulated 1,585 reads and 6 recommendations (English version only, the Spanish translation has 325 reads and 2 recommendations) so far, gaining a substantial acceptance. This initiative has been also highlighted by the \emph{Principality of Asturias} Science advisor \cite{Borja19} and in the \textit{Spanish Association for Science Advance (AEAC)} blog \cite{AEAC19}. Additionally, a description of this initiative has also been published in the online divulgation journal \textit{CompartiMOSS}, describing it from the student interaction point of view \cite{CompartiMOSS}.   

Future work will integrate this approach to reinforce problematic concepts in other courses related to computer security and administration \cite{RedondoASW}. These will complement other independent initiatives we plan to use, aimed to enhance learning from different points of view \cite{Varela18}, or with learning-focused tools \cite{Cuesta19}. Fanzines will also be used to improve the understanding of the most problematic theory and laboratory concepts of these courses identified by the student feedback, so the application of the rest of the initiatives could be more successful. Finally, we are also studying the feasibility of using an authoring tool (like in \cite{Zaibon2018EnhancingPO}) as an alternative to improve fanzine creation times.

\section*{Data Availability}

Zines described in this paper are available to the general public in \cite{RedondoZine20}. Ongoing future zine-related projects, related to cybersecurity, are also available in \cite{RedondoZine21}.

\section*{Acknowledgment}

The author of the paper wishes to thank researcher Daniel Gayo Avello (ORCID: 0000-0002-4705-6891) for the valuable feedback given to improve this paper.

\bibliographystyle{IEEEtran}
\IEEEtriggeratref{30}
\bibliography{jmredondo.zines.bib}


\end{document}